\documentclass[
aip,
jcp,
numerical,
reprint
]{revtex4-1}
\usepackage{graphicx}
\usepackage{float}
\usepackage{amsmath}
\newcommand{\bra}[1]{\left\langle \, #1 \, \right|}
\newcommand{\ket}[1]{\left| \, #1 \, \right\rangle}
\newcommand{\braket}[2]{\left\langle \, #1 \,|\, #2 \, \right\rangle}
\def\hT{{\hat T}}

\def\hW{{\hat W}}
\newcommand{\intra}{intra-system steepening }
\newcommand{\intraspace}{intra-system steepening}

\graphicspath{
{figures/pdf/}
}

\begin{document}
\title{Exact Maps in Density Functional Theory for Lattice Models}
\author{Tanja Dimitrov}
\email[Electronic address:\;]{dimitrov@fhi-berlin.mpg.de}
\affiliation{Fritz-Haber-Institut der Max-Planck-Gesellschaft, Faradayweg 4-6, D-14195 Berlin-Dahlem, Germany}
\affiliation{Max-Planck-Institut f\"ur Struktur und Dynamik der Materie, Luruper Chaussee 149, 22761 Hamburg, Germany}
\author{Heiko Appel}
\email[Electronic address:\;]{appel@fhi-berlin.mpg.de}
\affiliation{Fritz-Haber-Institut der Max-Planck-Gesellschaft, Faradayweg 4-6, D-14195 Berlin-Dahlem, Germany}
\affiliation{Max-Planck-Institut f\"ur Struktur und Dynamik der Materie, Luruper Chaussee 149, 22761 Hamburg, Germany}
\author{Johanna I. Fuks}
\email[Electronic address:\;]{johannafuks@gmail.com}
\affiliation{Department of Physics and Astronomy, Hunter College and the Graduate Center of the City University of New York, 695 Park Avenue, New York, New York 10065, USA}
\author{Angel Rubio}
\email[Electronic address:\;]{angel.rubio@mpsd.mpg.de}
\affiliation{Fritz-Haber-Institut der Max-Planck-Gesellschaft, Faradayweg 4-6, D-14195 Berlin-Dahlem, Germany}
\affiliation{Max-Planck-Institut f\"ur Struktur und Dynamik der Materie, Luruper Chaussee 149, 22761 Hamburg, Germany}
\date{\today}
%%%%%%%%%%%%%%%%%%%%%%%%%%%%%%%%%%%%%%%%%%%%%%%%%%%%%%%%%%%%%%%%%%
%                            Abstract                            %
%%%%%%%%%%%%%%%%%%%%%%%%%%%%%%%%%%%%%%%%%%%%%%%%%%%%%%%%%%%%%%%%%%
\begin{abstract}
In the present work, we employ exact diagonalization for model systems on a real-space lattice to explicitly construct the exact density-to-potential and for the first time the exact density-to-wavefunction map that underly the Hohenberg-Kohn theorem in density functional theory. Having the explicit wavefunction-to-density map at hand, we are able to construct arbitrary observables as functionals of the ground-state density. We analyze the density-to-potential map as the distance between the fragments of a system increases and the correlation in the system grows. We observe a feature that gradually develops in the density-to-potential map as well as in the density-to-wavefunction map. This feature is inherited by arbitrary expectation values as functional of the ground-state density. We explicitly show the excited-state energies, the excited-state densities, and the correlation entropy as functionals of the ground-state density. All of them show this exact feature that sharpens as the coupling of the fragments decreases and the correlation grows. We denominate this feature as {\it \intraspace}. We show that for fully decoupled subsystems the \intra transforms into the well-known inter-system derivative discontinuity. An important conclusion is that for e.g. charge transfer processes between localized fragments within the same system it is not 
the usual inter-system derivative discontinuity that is missing in common ground-state functionals, but rather the differentiable \intra that we illustrate in the present work.
\end{abstract}
\date{\today}
\maketitle
\section{Introduction}
Over the last decades ground-state density-functional theory (DFT) has become a mature tool in material science and quantum chemistry \cite{Baerends1997, Capelle2006, Ziegler1991, Kohn1996, Jones2015}. Provided that the exact exchange-correlation (xc) functional is known, DFT is a formally exact framework of the quantum many-body problem. In practice, the accuracy of observables in DFT highly depends on the choice of the approximate xc-functional. From the local density approximation (LDA) \cite{Kohn1965}, to the gradient expansions such as the generalized gradient approximations (GGAs), e.g. Perdew-Burke-Enzerhof (PBE) \cite{Perdew1996} and the hybrid functionals such as B3LYP \cite{Becke1993}, to the orbital-functionals such as optimized effective potentials \cite{Kuemmel2008} and to the range-separated hybrids such as HSE06 \cite{Heyd2003}, the last decades have seen great efforts and achievements in the development of functionals with more accurate and reliable prediction capability. \\
 Nonetheless, available approximate functionals such as the LDA, the GGA's and the hybrid functionals have known shortcomings to model gaps of semiconductors \cite{Hai2011}, molecular dissociation curves \cite{Caruso2013}, barriers of chemical reactions \cite{Cohen2012}, polarizabilities of molecular chains \cite{Champagne1998, Champagne1999}, and charge-transfer excitation energies, particularly between open-shell molecules \cite{Jacob2012}. \\ 
Recent advances in functional development such as optimally-tuned range separated functionals \cite{Baer2012}, ensemble density functional theory \cite{Kraisler2015, Kraisler2013} and local scaling corrections \cite{Cohen2015}, logarithmically enhanced factors in gradient approximations \cite{Armiento2013} and the particle-particle random-phase approximation \cite{Aggelen2013} can diminish or even cure some of the above mentioned shortcomings but not all of them. \\
Shortcomings of approximate functionals indicate that some important qualitative features of the exact functional are not (sufficiently well) captured. A common example is the delocalization error as in the case of stretched molecules, where approximate functionals such as LDA and GGA's tend to artificially spread out the ground-state electron density in space \cite{Cohen2008}. Since in DFT every observable is a functional of the ground-state density the delocalization error transmits into all observables as functional of the density and in particular to the ground-state energy functional. As a consequence most approximations for the ground-state energy as functional of the particle number $N$ are either concave or convex functions between integer $N$'s \cite{Cohen2008c, Cohen2015} and hence, violate the exact Perdew-Parr-Levy-Balduz condition \cite{Perdew1982} which states that the ground-state energy as a function of the particle number $E(N)$ is a linear function between integer $N$. The linearity of $E(N)$ leads to the commonly known derivative discontinuity \cite{Perdew1982} and is one exact condition on the xc-functional. Exact conditions on the xc-functional are a very useful tool in the development of new, improved functionals. In this paper we discuss an exact condition on the xc-functional that is relevant for systems consisting of well separated but mutually-interacting fragments, such as in stretched molecules. Among the approaches to model the limit of strongly correlated, low density systems with DFT we highlight the long range corrected hybrids \cite{Vydrov2006}, the generalization of the strictly correlated electron functional to fractional electron numbers \cite{Seidl1999b, Seidl2007, Gori2009, Mirtschink2013} and the recently introduced local scaling correction, which imposes the linearity condition to local regions of the system, correcting both energies and densities and affirming the relevance of modelling fractional electron distributions to reduce the delocalization error \cite{Cohen2015}.\\
Exactly solvable model systems have shown to provide useful insight essential to understand the failures of approximate xc-functionals and to develop new and improved approximations. For example, by studying one-dimensional model systems of few electrons it was shown that in the dissociation limit of molecules, the exact xc-potential as function of the spatial coordinate develops steps and peaks \cite{Buijse1989, Leeuwen1994, Gritsenko1994, Gritsenko1995, Gritsenko1996, Tempel2009, Helbig2009}. Such features are manifestations of strong-correlation and the absence of such features in approximate functionals results in delocalization errors. \\
Studies of exact ground-state xc-functionals for lattice models include the exact one-to-one map between ground-state densities and potentials computed for a half-filled one-dimensional Hubbard chain in Ref.~\cite{Capelle2003} using the Bethe Ansatz, for the one-site and double-site Hubbard models in full Fock space in Ref. \cite{Carrascal2012, Carrascal2015} and for the two-electron Hubbard dimer via constraint search in Ref.~\cite{Fuks2013}, among others. For such lattice models the Hohenberg-Kohn theorem \cite{Hohenberg1964} can be generalized by replacing the real-space potentials and densities by on-site potentials and on-site occupations \cite{Gunnarson1986, Schoenhammer1995}. The finite Hilbert space of lattice models permits the construction of the exact density-to-potential map. The question arises what can be learned about realistic three-dimensional systems by studying one-dimensional lattice models. Recently it was shown \cite{Fuks2014a, Fuks2014b} that the time-dependent exact xc-functional of the one-dimensional Hubbard dimer in the strongly-correlated limit develops the same step feature as the real-space one-dimensional model studied in Ref.~\cite{Fuks2013b}. Reference calculations of Ref.~\cite{Wagner2012} show that one-dimensional model systems capture the essence of three-dimensional systems when studying strong-correlation in DFT. \\
In this work, we study the exact density-to-potential and density-to-wavefunction map of a one-dimensional lattice model with a system size that still allows to exactly diagonalize the Hamiltonian in full Fock space. For different values of the external potential in the Hamiltonian we perform exact diagonalization of the Hamiltonian. Each diagonalization gives us all eigenfunctions and eigenenergies of the system, where the eigenstate with lowest eigenenergy corresponds to the ground state. We use the ground-state of each exact diagonalization corresponding to a fixed external and fixed chemical potential to construct both one-to-one maps, i.e.~the map between on-site potentials and ground-state on-site occupations (ground-state densities), and the map between ground-state densities and ground-state wave-functions. To illustrate the latter, we numerically construct the configuration-interaction (CI) coefficients of the wave-function expansion as functionals of the ground-state density. We study the exact features of these maps for systems with different ratio of discrete values of the kinetic hopping probability $\lambda_t$ to the electron-electron interaction strength $\lambda_w$. This allows us to study the exact maps from the non-interacting to the strictly-localized electron limit while we gradually change the correlation of the system. We illustrate how the distinctive features of the exact density-to-potential map transmit into the wavefunction-to-density map, and further into expectation values and transition matrix elements of {\it arbitrary} operators as functionals of the density.\\ 
We show that in approaching the limit of strongly correlated electrons, i.e. $\frac{\lambda_t}{\lambda_w}\rightarrow 0$, the gradient of the exact density-to-potential map steepens. We denote this feature as \intra which gradually builds up within the system as the hopping probability favoring the delocalization of electrons decreases and the electron-electron interaction favoring the localization increases. In the strictly localized electron limit, where $\lambda_t=0$, we see that the \intra transforms into the step-like inter-system derivative discontinuity. \\
We find that qualitative features such as the \intra and the inter-system derivative discontinuity of the density-to-potential map are already captured by a two-site lattice model. In the case of a two-site model, each site can be regarded as 
a subsystem. With increasing distance between the subsystems of the system, the hopping probability decreases and the localization of the electrons on each site increases. If the sites are infinitely apart, the subsystems are truly separated and the electrons are strictly localized on each site. We simulate the infinite separation in the two-site model by setting the hopping parameter in the kinetic operator $\lambda_t$ strictly to zero. Since the kinetic energy is strictly zero, this limit is the classical limit. However, setting $\lambda_t$ equal to zero allows us to imitate the infinite bond-stretching of the molecular model, where the distance of the molecular wells $d$ goes to $\infty$. In this limit, $\lambda_t=0$ implying $d\rightarrow \infty$, the \intra of the density-to-potential map becomes the standard step-like inter-system derivative discontinuity.\\ 
Arbitrary observables and transition-matrix elements are affected by the presence of the \intra and the inter-system derivative discontinuity, 
and in particular by the lack of it in approximate functionals. We illustrate how both features are 
transmitted to the ground- and excited-state 
energy, the excited- and transition-state density and to the correlation entropy functionals. \\ \\ 
The paper is organized as follows. In section \ref{sec:exact_map} we present the exact maps between local potentials, 
ground-state wavefunctions and ground-state densities. In section \ref{sec:model} we introduce the lattice model and the methodology that we employ in
the present work. Section \ref{sec:features} is dedicated to the study of the \intra in the exact density-to-potential map when approaching 
the strictly-localized limit (i.e. the strongly correlated limit) and its transition into a real inter-system derivative discontinuity for truly separated subsystems. 
In section \ref{sec:wf_and_observables} we use the potential-to-density map to construct the 
CI coefficients of the ground- and excited-state wavefunction expansions as
explicit functionals of the ground-state density. Ground-state degeneracies leave topological scars in the electron density \cite{Baer2010}. We illustrate how these degeneracies and furthermore also near-degeneracies of the eigenenergies of the system affect the ground- and excited-state expectation 
values and transition matrix elements of relevant operators as functionals of the ground-state density. Finally, in section \ref{sec:summary}, we summarize our findings and give an outlook for future work.
\section{Exact mappings}
\label{sec:exact_map}
\begin{figure}[b]
   \includegraphics[width=0.5\textwidth,natwidth=6.78in,natheight=4.32in]{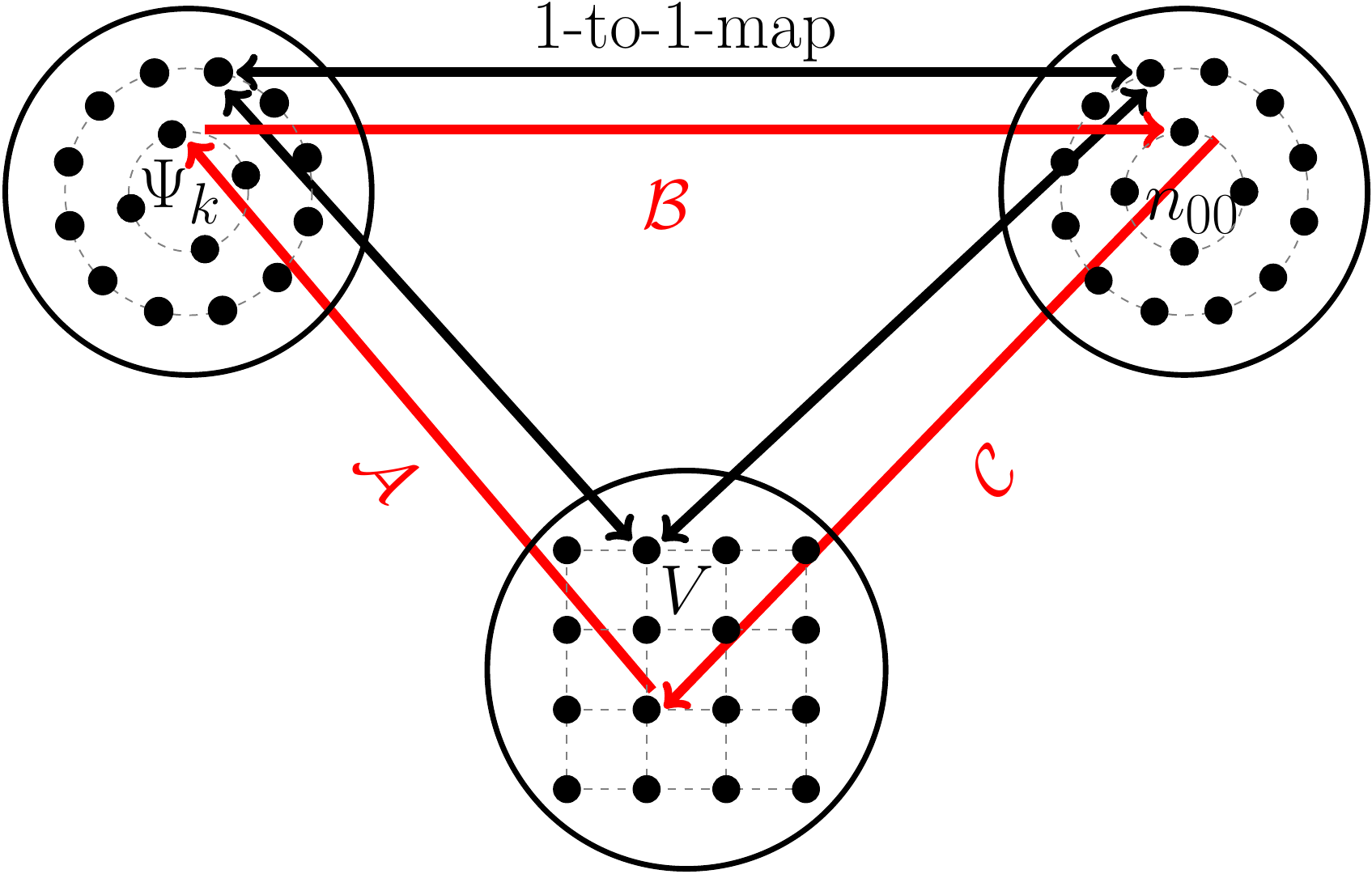}
   \caption{Schematic illustration of the exact mapping between $N$-electron wavefunctions 
         $\Psi_k$, local potentials $V$, and ground-state electron densities $n_{00}$. The maps are 
         depicted as red arrows, where $\mathcal{A}$ maps $V$ onto $\Psi_k$, $\mathcal{B}$ maps $\Psi_k$ onto $n_{00}$ 
         and $\mathcal{C}$ maps $n_{00}$ onto $V$. Black arrows indicate the bijectivity of each of these one-to-one maps. Note that every element in $V$ has an exact one-to-one 
         equivalent in $n_{00}$ and $\Psi_0$. }
    \label{fig:map_hk}
\end{figure}
To understand which features approximate functionals are missing, it is instructive to explicitly construct and to analyze the exact maps 
between the ground-state wavefunction $\Psi_0$, the local potential $V$, and the ground-state density 
{$n_{00}$, sketched in Fig.~\ref{fig:map_hk}. 
For fixed electron-electron interaction $\hat{W}$, the many-body Schr\"odinger equation, 
\begin{align}
(\hat{T}+\hat{W}+\hat{V}) \Psi_k = E_k \Psi_k \text{,}
\label{eq:SEQ}
\end{align}
defines a unique map between the set of local potentials $V$ and the set of energy eigenstates $\Psi_{k}$, 
depicted as map $\mathcal{A}$ in Fig.~\ref{fig:map_hk}. The ground-state  density  $n_{00}$ can be computed as usual according to 
\begin{align}
n_{00}(\vec{r})=N \int d\vec{r}_2...d\vec{r}_N |\Psi_0(\vec{r}_2..,\vec{r}_N)|^2\text{,}
\end{align}
which establishes a unique map 
from the set of $N$-electron ground-state wavefunctions $\Psi_0$ to the set of $N$-electron ground-state 
densities $n_{00}$. Hohenberg and Kohn \cite{Hohenberg1964} proved that the map $\mathcal{C}$ in Fig.~\ref{fig:map_hk} between $n_{00}$ and $V$
is one-to-one and unique if $V$-representability is fulfilled \cite{Kryachko1991, Kohn1983, Englisch1983B, Englisch1983, Chayes1985}. Assuming existence of this one-to-one density-to-potential map 
allows in principle to construct any ground-state observable as a unique functional of the ground-state density $n_{00}$,
\begin{equation}
 O_{00}[n_{00}]=\langle \Psi_0[n_{00}]| \hat{O}|\Psi_0[n_{00}]\rangle\text{.}
\end{equation}
Note that as a consequence of the one-to-one $V$-to-$n_{00}$ map, 
the Schr\"odinger equation additionally establishes a one-to-one map between $n_{00}$ and the excited-state 
wavefunctions $\Psi_{k}, k \neq 0$.
As a consequence, excited-state expectation values with $k=l>0$, and transition matrix elements with $k\neq l$,
 can be computed as functionals of the ground-state density using
\begin{equation}
 O_{{kl}}[n_{00}]=\langle \Psi_k[n_{00}]| \hat{O}|\Psi_l[n_{00}]\rangle\text{.}
 \label{eq:transition_matrix_elements}
\end{equation}
The ground-state energy $E_0$ and the ground-state density $n_{00}$ can be accessed
using the variational principle,
\begin{equation}
 E_0 = \min_{n} E_{V}[n], \quad E_0 < E_{V}[n], \quad n\neq n_{00}.
\label{eq:hkvp1}
\end{equation}
Given an external potential $V$ the total energy is computed as $E_{V}[n]=F_{HK}[n] + \int n(\vec{r}) V(\vec{r}) d^3{r}$. 
In the Levy-Lieb constrained search formulation \cite{Levy1982, Lieb1983} the Hohenberg-Kohn energy functional $F_{HK}$ is found as the minimum
over all possible $N$-electron densities $n$, of the expectation value of kinetic plus electron-electron interaction operator
\begin{eqnarray}
F_{\text{HK}}[n]=\underset{\Psi\rightarrow n}{\text{min}}\bra{\Psi[n]} \hT+\hW \ket{\Psi[n]}\text{.}
\label{eq:HK}
\end{eqnarray}
In the following, we illustrate the features of the exact density-to-potential and density-to-wavefunction maps 
explicitly for our model systems.
\section{Lattice Model}
\label{sec:model}
In the present work, we restrict ourselves to one-dimensional lattice systems for which the construction of exact functionals via exact diagonalization is numerically feasible. On a lattice, the potential becomes an on-site potential $V(x) \to v(x_i)$, the density
a site-occupation, $n(x) \to n(x_i)$, and the integral becomes a sum over sites $i$, $\int dx \to \sum_i$\cite{Gunnarson1986}. Furthermore, the kinetic energy operator becomes a nearest-neighbor hopping term.
\subsection{Lattice Hamiltonian}
For $N$ interacting electrons in one spatial dimension we consider Hamiltonians of the form
\begin{equation}
      \hat{H}^\varphi =\hat{H}(\varphi) =  \lambda_t(\varphi)\hat{T}+ \lambda_w(\varphi)\hat{W} + \hat{V} + \mu \hat{N} \text{,}
        \label{eq:hamiltonian_fs}
\end{equation}
where the parameter $\mu$, connected to the particle number operator $\hat{N}$, acts as a Lagrange multiplier shifting the
state with lowest energy to blocks with different particle number $N$ in Fock space. To switch between different coupling limits,
we introduce the amplitude of the kinetic hopping $\lambda_{\text{t}} = r \cos(\varphi)$ and the strength of the electron-electron interaction $\lambda_{\text{w}} = r \sin(\varphi)$ as parameters in polar representation with radius $r$ and angle $\varphi$, see Fig.~\ref{fig:map_varphi}. The 
limit $\lambda_w \rightarrow 0$, i.e. $\varphi=0$, corresponds to non-interacting electrons and the limit $\lambda_t \rightarrow 0$, i.e. $\varphi=\frac{\pi}{2}$,
to site-localized electrons. Without loss of generality we choose $r=\sqrt{2}$ and $\varphi \in [0,\frac{\pi}{2}]$. 
Throughout this work we use atomic units $\hbar=m=e=1$.
\begin{figure}[t]
\includegraphics[width=0.3\textwidth,natwidth=2.19in,natheight=2.11in]{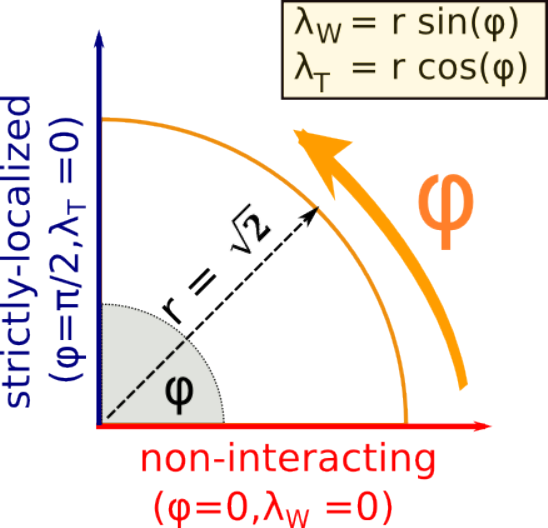}
\caption{Illustration of the kinetic hopping probability $\lambda_{\text{t}}$ and the strength of electron-electron interaction $\lambda_{\text{w}}$ in polar representation. }
\label{fig:map_varphi}
\end{figure}
\\
Next, we introduce the operators of the lattice model with $M$ sites and lattice spacing $dx$. 
In second-order finite difference representation the kinetic energy operator with nearest neighbor hopping $t_0=\frac{1}{dx^2}$ reads
\begin{eqnarray}
        \hat{T}&=& -\frac{t_0}{2}  \sum_{l=1}^{M}\sum_{\sigma}  \hat{c}_{l, \sigma}^{\dag}\hat{c}_{l+1, \sigma} + \hat{c}_{l+1, \sigma}^{\dag}\hat{c}_{l, \sigma}
        - 2 \hat{c}^{\dag}_{l, \sigma} \hat{c}_{l, \sigma} \text{,}
\label{eq:kinetic_operator_fs}
\end{eqnarray}
where $\hat{c}^{\dagger}_{l, \sigma}$ and $\hat{c}_{l, \sigma}$ denote creation and annihilation operators 
of an electron placed on site $l$ with spin projection onto the z-axis $\sigma$. 
Usually, the hopping $t_0$ changes with the lattice spacing $dx$. However, we choose to leave $dx$ fixed and use the parameter $\lambda_t$ and $\lambda_w$ instead. The last term in Eq.~\ref{eq:kinetic_operator_fs} corresponds to on-site hopping. For model systems this term is usually not taken into account. Here
we keep the term to allow for a consistent first and second quantized treatment of the Hamiltonian.
We study the non-local soft-Coulomb electron-electron interaction,
\begin{equation}
        \hat{W}_{\text{SC}}=\sum_{l, m, \sigma, \sigma'} \frac{\hat{c}^{\dag}_{l, \sigma}\hat{c}^{\dag}_{m, \sigma'}\hat{c}_{m, \sigma'}\hat c_{l, \sigma} }{2\sqrt{(dx(l-m))^{2}+a}}\text{,}
\label{eq:soft_coulomb}
\end{equation}
acting on particles located at sites $l$ and $m$ with spins $\sigma$ and $\sigma'$. Throughout this work the Coulomb 
interaction is softened by the parameter $a = 1$. The external potential
\begin{equation}
        \hat{V}=\sum_{l=1}^M  v_{l}\hat{c}^{\dag}_{l}\hat{c}_{l}\text{,}
	\label{eq:external_potential}
\end{equation}
introduces a potential difference between the sites in the lattice, which depending on its strength, shifts 
the electron density among the sites in the lattice. 
We restrict ourselves to two different scenarios for which exact diagonalization is still possible, similar to Ref.~\cite{Sanchez2014, Cohen2014}. In case (i) we
consider two spin-singlet electrons on $M=206$ sites. The particles are confined in a box from $x=-10.25$ a.u. to $x=+10.25$ a.u.
with zero boundary conditions and a lattice spacing of $dx = 0.1$ a.u. and $\lambda_t=\lambda_w=1$. 
To mimic the bond-stretching in molecular systems, we consider an external potential 
\begin{align}
v_l =  \frac{Z_1(\alpha)}{\sqrt{(x_l-\frac{d}{2})^2+1}} + \frac{Z_2(\alpha)}{\sqrt{(x_l+\frac{d}{2})^2+1}}+\frac{Z_1(\alpha)Z_2(\alpha)}{\sqrt{(d^2+1)}},\\ 
 Z_1(\alpha) = -\alpha,\;\; Z_2(\alpha) = -(2-\alpha)
\label{eq:VZ1}
 \end{align}
with two atomic wells separated by distances ranging from $d=2$ to $d=8$ a.u..
The depth of the wells is given by the nuclear charges $Z_1$ and $Z_2$ which we modulate with the parameter $\alpha\in[0,2]$.
We will see that the essence of such a system is already captured by a two-site model. As case (ii), we
consider $M=2$ sites in the lattice with a distance $dx=\frac{1}{\sqrt{2}}$, where we vary the parameters $\lambda_t$ and $\lambda_w$. In this case, the system size allows to perform exact diagonalization in the full Fock space of the model.
\subsection{Methodology}
\label{subsec:exact_diagonalization}
To explicitly construct the one-to-one map between external potentials and ground-state densities, we diagonalize 
the Hamiltonian introduced in Eq.~(\ref{eq:hamiltonian_fs}) for different external potentials $v_m$, but fixed $\varphi$. The external potential take values $v_m = m \Delta_v$, 
where $\Delta_{v}$ is the numerical step size, and $m$ is the step numbers. 
For completeness, although not shown in the present work, similarly the Hamiltonian can be diagonalized for different chemical 
potentials $\mu_k$ with the chemical potential values $\mu_k = k \Delta_{\mu}$, the numerical step size $\Delta_{\mu}$ and the step number $k$. In this way all functionals are constructible as function of the particle number $N$ and can be studied in complete Fock-space. Here we fix the chemical potential and select a discrete and uniformly distributed 
set of potentials from the continuous set V of possible external potentials in Fig.~\ref{fig:map_hk}.
In the next step we use exact diagonalization to compute the ground-state wavefunction $\Psi_0^\varphi$ and 
energy $E_0^{\varphi}$ corresponding to each value of $v_i$ and $\mu_i$ (but fixed $\varphi$). For each 
ground-state wavefunction, we compute the corresponding on-site ground-state density 
\begin{equation}
       n_{00}^\varphi(x_j)=\bra{\Psi_0^\varphi}\hat{n}(x_j) \ket{\Psi_0^\varphi}\text{,}
\end{equation}
where $j$ is the site subindex, and the spin-summed density operator reads
\begin{equation}
\hat{n}(x_j)=\hat{c}^{\dagger}_{j,\uparrow}\hat{c}_{j,\uparrow}+\hat{c}^{\dagger}_{j,\downarrow}\hat{c}_{j,\downarrow}\text{.}
\end{equation}
In addition to the ground-state wavefunction, the exact diagonalization gives us access to the excited-state 
wavefunctions $\Psi_{k\neq 0}^\varphi$, which allows us to compute excited-state observables and transition matrix elements of operators as functionals of
the ground-state density according to Eq.~ \ref{eq:transition_matrix_elements}. On a lattice with $M$ sites, the continuous one-dimensional ground-state density $n_{00}(x)$ becomes a vector $(n_{00}(x_1),n_{00}(x_2), ...,n_{00}(x_M))$. Hence, expectation values and transition matrix elements as functionals of the density become rank $M$ tensors
\begin{align}
O_{kl}^{\varphi}(n_{00}(x_1),...,n_{00}(x_M))=\nonumber\\
\langle\Psi_k^{\varphi}(n_{00}(x_1),...,n_{00}(x_M))|\hat{O}|\Psi_l^{\varphi}(n_{00}(x_1),...,n_{00}(x_M))\rangle \text{.}
\end{align}
In case (ii) where two sites in the lattice are considered, all functionals depend on the on-site densities  $n_{00}(x_1)$ and $n_{00}(x_2)$, i.e.~$\ket{\Psi_k^\varphi[n_{00}(x_1), n_{00}(x_2)]}$ and
$V^\varphi[n_{00}(x_1), n_{00}(x_2)]$. Instead of expressing all functional dependencies in terms of the variables $n_{00}(x_1)$ and $n_{00}(x_2)$, 
we rotate the coordinate system to the total particle number $N = n_{00}(x_1) + n_{00}(x_2)$ and the occupation difference 
$\delta n_{00}=n_{00}(x_1) - n_{00}(x_2)$ between the sites \cite{Fuks2013}.
To illustrate the wavefunction-to-density map, we expand the ground $(k=0)$ and the excited $(k>0)$ eigenstates $\ket{\Psi_k^\varphi}$ of the system in a complete set of Slater determinants $\ket{\Phi_q}$, 
\begin{equation}
\ket{\Psi_k^\varphi[\delta n_{00}, N]} = \sum_q \alpha_q^{\varphi,k}[\delta n_{00}, N] \ket{\Phi_q}\text{,}
\end{equation}
where we have chosen $\ket{\Phi_q}$ to be the eigenstates of the kinetic operator $\hat{T}$. This gives rise to the CI coefficients 
\begin{equation}
\alpha_q^{\varphi,k}[\delta n_{00}, N]=\braket{\Phi_q}{\Psi_{k}^\varphi[\delta n_{00}, N]}.
\label{eq:ci_coefficients}
\end{equation}
By writing  the CI-coefficients $\alpha_q^{\varphi,k}[\delta n_{00}, N]$ as explicit functionals of $\delta n_{00}$, 
we gain access to all ground- and excited-state expectation values or transition matrix elements of any operator, i.e.
\begin{align}
O_{kl}^\varphi [\delta n_{00}, N]=\bra{\Psi_k^\varphi[\delta n_{00}, N]}\hat{O}\ket{\Psi_l^\varphi[\delta n_{00}, N]}\nonumber\\
  = \sum_{q}\sum_{q'}\alpha_{q'}^{\varphi k \ast}[\delta n_{00},  N]\alpha_{q}^{\varphi l}[\delta n_{00}, N]\bra{\Phi_{q'}}\hat{O}\ket{\Phi_q} \text{.}
\label{eq:observables_ci}
\end{align}
A prime example is the Hohenberg-Kohn energy functional defined in Eq.~\ref{eq:HK}, which is the expectation value of $\hat{H}_{v_l=0}^{\varphi}=\lambda_{t}(\varphi)\hat{T}+\lambda_{w}(\varphi)\hat{W}$, i.e.
\begin{align}
F_{00}^\varphi[\delta n_{00}, N] =\bra{\Psi_0^\varphi[\delta n_{00}, N]}\hat{H}_{v_l=0}^{\varphi}\ket{\Psi_0^\varphi[\delta n_{00}, N]}\nonumber \\
  = \sum_{q,q'}\alpha_{q'}^{\ast}[\delta n_{00}, N]\alpha_{q}[\delta n_{00}, N]\bra{\Phi_{q'}}\hat{H}_{v_l=0}^{\varphi}\ket{\Phi_q}\text{.}
\label{eq:hk_ci}
\end{align}
For the two-particle singlet states, we compute the Hohenberg-Kohn functional for different values of $\varphi \in [0,\frac{\pi}{2}]$. Note the explicit dependence of $F_{00}^\varphi[\delta n_{00}, N]$ on the angle $\varphi$, since the Hohenberg-Kohn proof can only be established for fixed and given kinetic energy and particle-particle interaction. By changing the angle $\varphi$, we construct the exact energy functional $F_{00}^\varphi[\delta n_{00}, N]$ for different 
electron-electron interactions and kinetic terms, where
$\varphi=0$ is the non-interacting and $\varphi=\frac{\pi}{2}$ the infinitely correlated limit. 
Although the Hohenberg-Kohn ground-state energy functional is a very important example, our approach allows to construct
the exact density functionals for any observable of interest. We illustrate this for a few selected examples in the following
sections. Also we emphasize that throughout this work all functionals are constructed in the zero-temperature limit.
\section{Features of the exact Density-to-Potential Map}
\label{sec:features}
\begin{figure*}[t]
\includegraphics[width=\textwidth,natwidth=6.69in,natheight=1.94in]{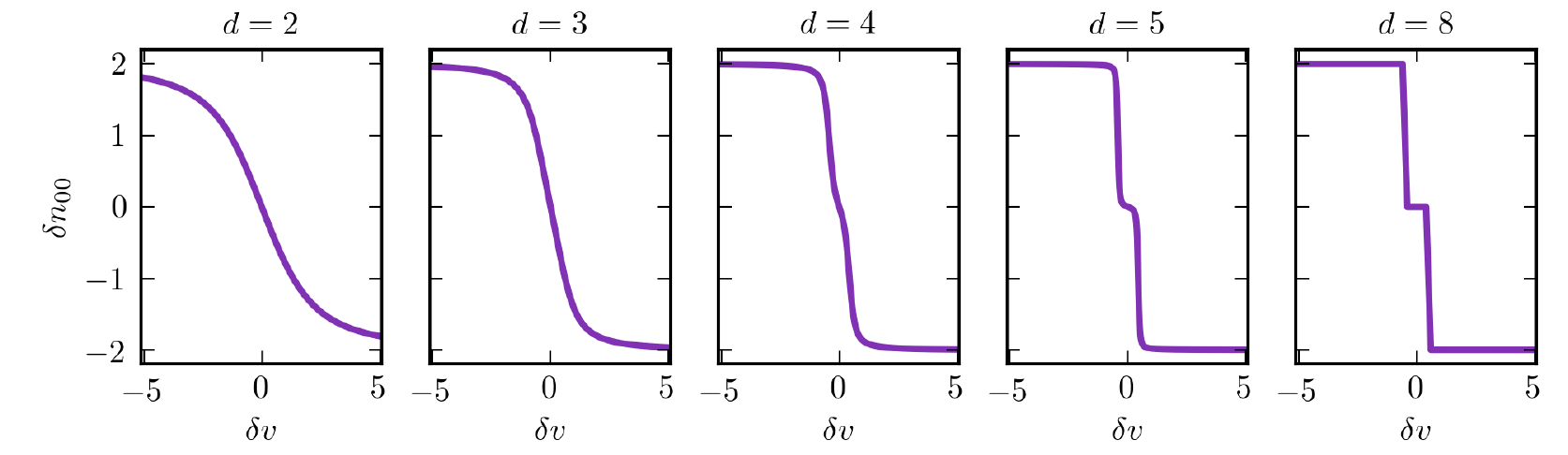}
\caption{Exact density-to-potential map for a one-dimensional diatomic molecule with nuclear charges 
$Z_1$ and $Z_2$, where we vary the potential difference $\delta v = Z_1-Z_2$ from $-5$ to $5$ for different atomic separations $d=2-8$ a.u. The density 
difference $\delta n_{00}$ corresponds to the electronic density summed over the left half-space minus the 
density summed over the right half-space as defined in Eq.~\ref{eq:dn_for_molecule}.
 The graph illustrates the influence of electron localization 
on the ground-state density-to-potential map. From left to right the distance of the molecular wells 
increases while the gradient of the density-to-potential steepens with increasing distance $d$  and hence, decreasing coupling of the 
fragments of the system. We denote this feature of the density-to-potential map as \intra (see text for details).}
\label{fig:exact-map-dn-dv-molecule}
\end{figure*}
\begin{figure*}[t]
\includegraphics[width=\textwidth,natwidth=6.51in,natheight=5.35in]{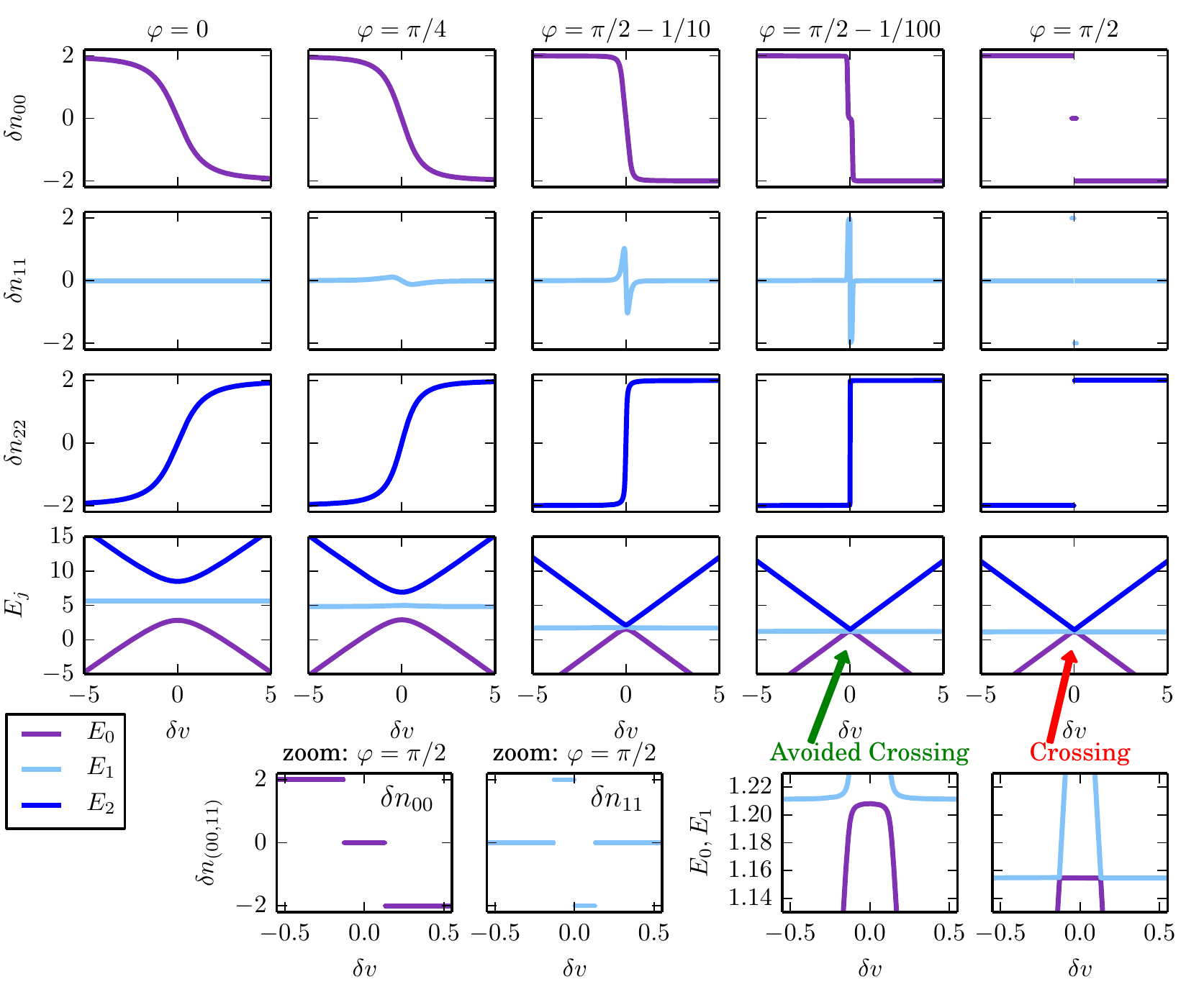}
\caption{Exact density-to-potential map for a two-site lattice model using soft-Coulomb interaction. Despite its reduced dimensionality essential features of the density-to-potential map of the molecular model system are already captured by a two-site model, as can be seen by comparing Fig.~\ref{fig:exact-map-dn-dv-molecule} and Fig.~\ref{fig:exact-map-dn-dv}. The graphs illustrate how the electron localization is captured in the ground- and excited-state density-to-potential maps and in the eigenenergies. Upper panel: exact ground-state density as function of the external potential, i.e. $\delta n_{00}(\delta v)$. Second panel: exact first excited-state density 
 as function of the external potential, i.e. $\delta n_{11}(\delta v)$. Third panel: exact second excited-state density 
 as functional of the external potential, i.e. $\delta n_{22}(\delta v)$. Lower panel: eigenenergies of the two-particle 
singlet states as functional of the external potential $E_j(\delta v)$, where $E_j$ corresponds to the eigenstate $\ket{\Psi_j}$ and to the density differences $\delta n_{jj}=\bra{\Psi_j}\delta \hat{n}\ket{\Psi_j}$.
Inset at the bottom on the left-hand side: Detailed view of the ground-state and the first excited-state density functionals $\delta n_{00}(\delta v)$ and $\delta n_{11}(\delta v)$ in the strictly localized limit. Inset at the bottom at the right-hand side: avoided and real crossings of eigenenergies. From left to right the angle $\varphi$ increases the correlation in the system going from the non-interacting ($\varphi=0$)
to the strictly-site-localized electron limit ($\varphi=\frac{\pi}{2}$). In the molecular model system of Fig.~\ref{fig:exact-map-dn-dv-molecule} this corresponds 
to an increasing distance $d$ of the molecular wells. The gradient of all three densities steepens whenever the 
corresponding eigenstate as functional of the external potential comes close to an avoided crossing. We denote this exact feature of the density-to-potential map as \intraspace. In the strictly localized limit ($\varphi=\frac{\pi}{2}$) the \intra transitions into the inter-system 
derivative discontinuity while the avoided crossing transitions into a real-crossing with degenerate eigenenergies.} 
\label{fig:exact-map-dn-dv}
\end{figure*}
We start our analysis for the Hamiltonian of case (i), where we consider a diatomic molecule with different
interatomic separations. While the full density-to-potential map is a high-dimensional function for
$M=206$ sites and impractical to visualize, the essence of the bond stretching can be captured by the integrated 
densities of fragments of the system. A natural choice to partition the system into its fragments, is to divide the total molecular charge distribution at its minima into different Bader basins \cite{Bader1970, Henkelman2006}. By integrating the density over each of these Bader basins, the high dimensionality of the density in real-space reduces drastically. For our diatomic model the partitioning reduces the dimensionality from 206 to two, by mapping the sites in the grid onto the basins. We can then refer to each basin as a
{\it effective site} in real space and regard the density difference between the basins as density difference 
between the two sites. 
For the simple diatomic molecule in one dimension, we simply divide the system in two 
equal half-spaces, and construct the density-difference according to 
\begin{align}
\delta n_{00}=\sum_{i=1}^{M/2}n_{00}(x_i)-\sum_{i=M/2+1}^{M}n_{00}(x_i)\text{.}
\label{eq:dn_for_molecule}
\end{align}
To obtain the potential difference between the two basins, we take the difference between the maximum depth of the molecular potential wells of each basin and define the potential difference as $\delta v = Z_1(\alpha) - Z_2(\alpha)$. Note, the potential difference can be tuned by changing the nuclear charge of the two atoms continuously with the parameter $\alpha$ of Eq.~\ref{eq:VZ1}. The resulting {\it effective} density-to-potential map for our diatomic model is shown in Fig.~\ref{fig:exact-map-dn-dv-molecule}. 
Starting from left to right we increase the distance between the molecular wells. The effective density-to-potential
map starts out with a smooth monotonic shape. When the distance of the atoms is increased the gradient of the 
density-to-potential map steepens, leading ultimately to steps in the density values for the infinitely 
separated limit. \\
The very same qualitative behavior can be found for a simple two-site lattice system. As a second example
we consider therefore the Hamiltonian of case (ii). In this case we construct the exact density-to-potential 
map for the two-particle singlet states of the two-site lattice model. The
results are shown in Fig.~\ref{fig:exact-map-dn-dv}. In addition to the ground-state density-to-potential 
map in the first row of Fig.~\ref{fig:exact-map-dn-dv}, the second and third row show the first and 
second excited-state density-to-potential map, and the fourth row shows the eigenenergies $E_0$, $E_1$ and $E_2$  as function of the external potential difference between the two sites in lattice. From left to right, $\varphi$ increases, i.e. the electron-electron 
interaction favoring the localization of the electrons increases, whereas the kinetic energy favoring the 
delocalization decreases, i.e. $\frac{\lambda_t}{\lambda_w} \rightarrow 0$. This localization is reflected by the steep gradient of the ground- and excited-state 
densities $\delta n_{00}$, $\delta n_{11}$ and $\delta n_{22}$ as function of the external potential difference $\delta v = v_1-v_2$.
For the potential difference we select values from -5 to 5, shifting the electron density from one site in the lattice to the other.
Setting $\varphi=0$ in Eq.~(\ref{eq:hamiltonian_fs}) corresponds to non-interacting electrons, where the 
eigenfunctions are single-particle Slater-determinants. In this limit the density-to-potential map for our model
can be found analytically
\begin{align}
\delta n_{00}^{\varphi=0}(\delta v, \lambda_t, dx)=- \frac{4 dx^2 \delta v}{\sqrt{ 4 dx^4 \delta v^2+\lambda_t^2(\varphi=0) }}.
\end{align}
The map behaves smoothly as can be seen in the leftmost figure in the first row of Fig.~\ref{fig:exact-map-dn-dv}.\\
Approaching the strictly-localized limit, i.e. $\varphi \to \frac{\pi}{2}$, the slope of the exact 
density-to-potential map sharpens until the map develops a characteristic feature, which we denote as \intraspace. 
The \intra of the  gradient of the density-to-potential map corresponds to the localization of the electrons in the respective subsystems. Near the strictly-localized limit, e.g. $\varphi=\frac{\pi}{2}-\frac{1}{100}$, the electrons are highly-localized 
on the sites.\\
In the strictly-localized electron limit $\varphi=\frac{\pi}{2}$ the hopping parameter is equal to zero. In this
limit the system 'breaks' into two physical disconnected sites of integer occupation, the Hamiltonian reduces to $\hat{H}^{\varphi=\frac{\pi}{2}}=\hat{W}+\hat{V}$ and hence commutes with the position operator $\hat{x}=\sum_{m=1}^{M}\sum_{\sigma}x_{m}\hat{c}_{m,\sigma}^{\dagger}\hat{c}_{m,\sigma}$ with $x_{m}=-(M+1)/2 dx + m dx$. 
\begin{align}
	[\hat{H}^{\varphi=\frac{\pi}{2}},\hat{x}]=0\text{,}
\end{align}
and the eigenfunctions of $\hat{H}$ 
are diagonal in the eigenbasis of the position operator. The three two-particle singlet states correspond to the physical situations where both electrons are located on 
site one, i.e. $\ket{\Psi_0^{\varphi=\frac{\pi}{2}}[\delta n_{00}=+2]}=\hat{c}^{\dagger}_{1\downarrow}\hat{c}^{\dagger}_{1\uparrow}\ket{0}$, both 
electrons are on site two, i.e. $\ket{\Psi_0^{\varphi=\frac{\pi}{2}}[\delta n_{00}=-2]}=\hat{c}^{\dagger}_{2\downarrow}\hat{c}^{\dagger}_{2\uparrow}\ket{0}$, or where the electrons are delocalized over both sites, i.e. 
$\ket{\Psi_0^{\varphi=\frac{\pi}{2}}[\delta n_{00}=0]}=\frac{1}{\sqrt{2}}\left(\hat{c}^{\dagger}_{1\downarrow}\hat{c}^{\dagger}_{2\uparrow} - \hat{c}^{\dagger}_{1\uparrow}\hat{c}^{\dagger}_{2\downarrow}\right)\ket{0}$. 
Depending on the ratio between the external potential difference $\delta v$ and the electron-electron repulsion strength $\lambda_w$, 
one of these three eigenstates is energetically more favorable and becomes the ground-state of the electronic system, see 
lower panel of Fig.~\ref{fig:exact-map-dn-dv}.
Using the strictly-localized ground-state wavefunction, the density difference $\delta n_{00}$ 
transitions from a continuous variable to a {\it discrete set} of integer values. Namely, 
the only possible values for the ground-state density differences are the integer values 
\[
   \delta n_{00}^{\varphi=\frac{\pi}{2}} (\delta v) = 
\begin{cases}
  -2, \\
   0, \\
  +2. 
\end{cases}
\]
In this limit different values of the external potential lead to the same density difference $\delta n_{00}$ 
as can be seen in the map for $\varphi=\frac{\pi}{2}$ in Fig.~\ref{fig:exact-map-dn-dv}. Therefore, the one-to-one map between 
$\delta n_{00}$  and $\delta v$ breaks down and the \intra transitions into the inter-system derivative discontinuity, since the two sites decouple and are becoming two separate systems.  Functionals in the distributional limit are a linear combination of the functionals of the degenerate densities as has been shown for the ground-state energy functional as functional of the particle number\cite{Perdew1982, Yang2000}.
Therefore, we connect the distributional points 
for all functionals via straight lines, i.e. $\delta n_{00} = \pm 2 (1-\omega) $ and $0\leq \omega \leq1$. In a physical 
picture each one of the disconnected sites can be seen as a system infinitesimally weakly connected to a grand-canonical particle reservoir. \\
Contrary to the widely discussed inter-system derivative discontinuity, which describes the piece-wise linear behavior of 
the energy as a function of the particle number $E[N]$, the \intra  describes the smooth behavior of the energy 
as functional of the density difference between fragments {\it within} the system $E[\delta n_{00}]$. Both features already show 
up in the density-to-potential map and transmit to all observables. The Hohenberg-Kohn energy functional is therefore 
only one specific example for the appearance of inter-system derivative discontinuity and \intraspace. The smooth behavior of the \intra is a consequence of the mixing of different quantum eigenstates around
avoided crossings, and the steps related to the inter-system derivative discontinuity directly result from intersections of eigenenergies, thus real crossings, 
see lower panel and inset of Fig.~\ref{fig:exact-map-dn-dv}. The inter-system derivative discontinuity appears when electrons are strictly-localized 
in states with different particle number. 
Note that  
the steepening of the 
gradient for $\delta n_{00}$ as well as for $\delta n_{11}$ and $\delta n_{22}$
arises whenever the eigenvalues of the Hamiltonian in  Eq.~(\ref{eq:hamiltonian_fs}) as function of 
the external potential become nearly degenerate. The connection between the avoided crossing and the steepening of the gradients functional is closely related to the finding of Ref. \cite{Tempel2009}, i.e. that the step feature of the exact xc-potential in space arises in the vicinity of the avoided crossing, when the bonding and antibonding orbitals become nearly degenerate. Without this step feature (and the peaks) of the exact xc-potential, the non-interacting electron density would artificially smear out over both basins and lack the \intra of the exact electron density-to-potential map.
For $\varphi=0$ all eigenvalues are non-degenerate, hence the density-to-potential map of all 
eigenstates behaves smoothly. When we approach the strongly-correlated limit at $\varphi \to \frac{\pi}{2}$,
the first and second excited-state energies approach each other $E_1[\delta v] \to E_2[\delta v]$ and
for $\varphi=\frac{\pi}{2}$ they become degenerate for $\delta v=0$, i.e. $E_1[\delta v] = E_2[\delta v]$ (see inset Fig.~\ref{fig:exact-map-dn-dv}).
Caused by a real crossing of the eigenenergies in the strictly-localized limit, the one-to-one correspondence with an external potential breaks down for all densities, i.e.
the ground-state and the excited-state densities. The density-to-potential map becomes a distribution in this limit and the Hohenberg-Kohn theorem doesn't apply.
\section{Features of the exact Density-to-Wavefunction Map}
\label{sec:wf_and_observables}
\begin{figure*}[ht]
\includegraphics[width=\textwidth,natwidth=6.69in,natheight=4.13in]{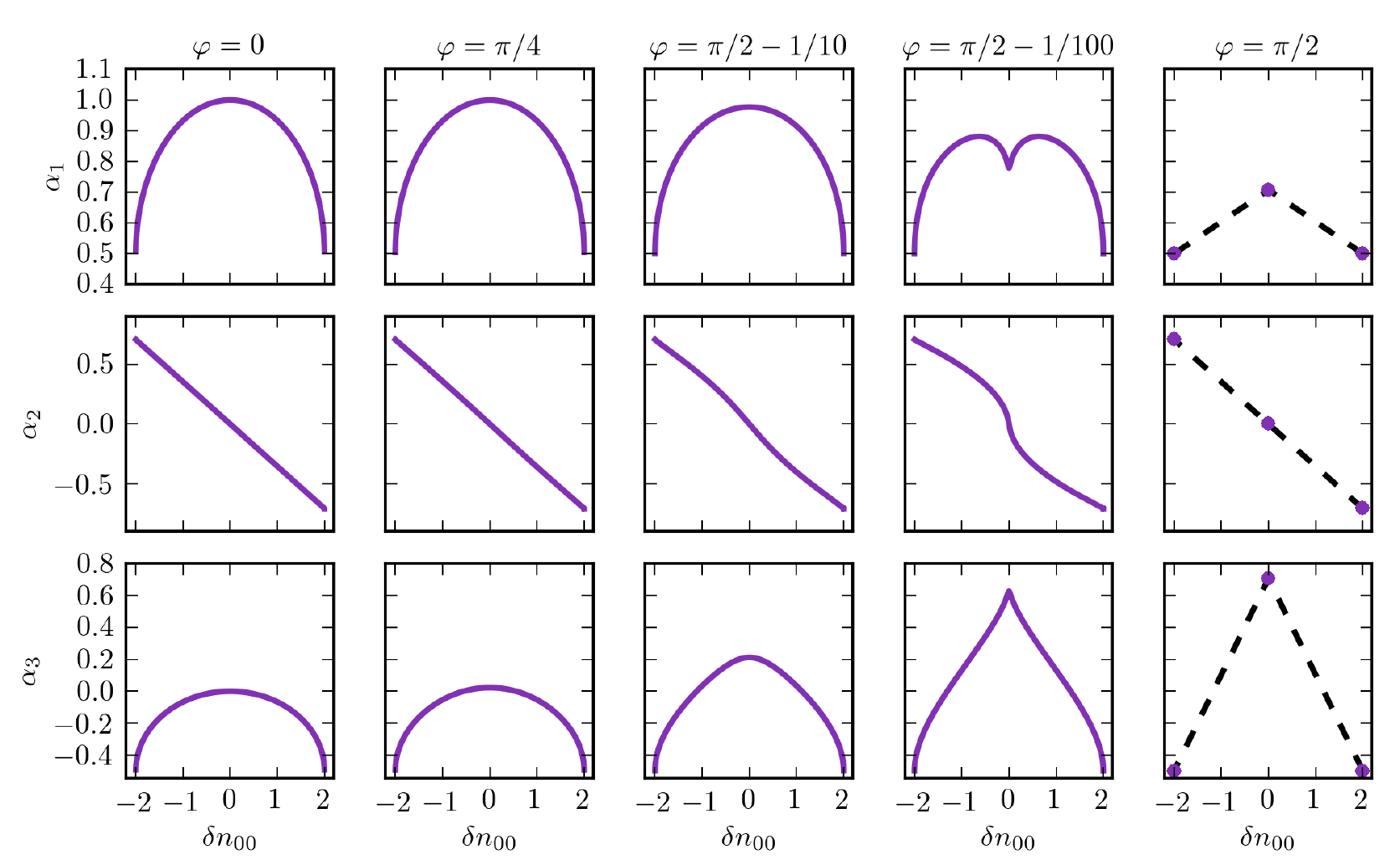}
\caption{CI coefficients of the two-particle ground-state wavefunction 
in the kinetic operator basis. From left-to-right we approach the strictly-localized limit ($\varphi=\frac{\pi}{2}$) and the gradient of all three CI coefficients steepens. 
For $\varphi=\frac{\pi}{2}$ the CI coefficients take only discrete values which can be interpolated linearly (dashed lines) due to the degeneracy of the eigenstates in the strictly localized limit.}
\label{fig:exact-map-alpha-dn}
\end{figure*}
The inter-system derivative discontinuity and the \intra discussed in the previous section are exact properties of the density-to-potential map. As a consequence, also the exact wavefunction and hence, all exact observables - here, in particular the ground-state Hohenberg-Kohn energy functional- as function of the exact density inherit the \intra and the inter-system derivative discontinuity. In the following sections we illustrate this fact. In particular, we 
show how these features show up in the CI-coefficients, and consequently in the energy, the excited-densities and in the correlation entropy functional. 
\subsection{Exact Configuration Interaction Coefficients as Functionals of the Ground-State Density}
\label{subsec:ci_coefficients}
To construct the density-to-wavefunction map, we expand the correlated ground- and excited-state wavefunctions from the
exact diagonalization of the Hamiltonian in a complete set of Slater determinants $\ket{\Phi_q}$. This gives rise to CI coefficients as functionals of the
ground-state density as defined in Eq.~\ref{eq:ci_coefficients}. Clearly, each choice for the set of Slater determinants $\ket{\Phi_q}$ induces a different set of CI functionals. Here we choose as basis
set the determinants which are eigenfunctions of the free kinetic energy operator. More specifically, we project the two-particle singlet ground-state wavefunction of the Hamiltonian in Eq.~\ref{eq:hamiltonian_fs} onto the three two-particle singlet eigenstates of the kinetic operator $\hat{T}$ to construct one of these sets for each different $\varphi$. The results are summarized in Fig.~\ref{fig:exact-map-alpha-dn}. Each row in the figure displays one of the ground-state CI coefficients 
as function of the density difference between the sites, $\alpha_q[\delta n_{00}]=\braket{\Phi_q}{\Psi_0[N=2,S^2=0,S_z=0,\delta n_{00}]}$.
For non-interacting electrons, the CI coefficients can be evaluated analytically. In our chosen basis the coefficients have no direct dependency on $\lambda_t$, 
\begin{align}
\alpha_1^{\varphi=0}\left[\delta n_{00}\right] &=-\frac{\left(\delta n^{2}_{00}-2\left(2+\sqrt{4-\delta n^{2}_{00}}\right)\right)\left(2+|\delta n_{00}| \right)}{4 \sqrt{-(-4+\delta n_{00}^2)(4+\delta n_{00}^2+4|\delta n_{00}|)}}\\
\alpha_2^{\varphi=0}\left[\delta n_{00}\right] &=-\frac{\delta n_{00}}{2\sqrt{2}}\\
\alpha_3^{\varphi=0}\left[\delta n_{00}\right] &=-\frac{\left(-4+\delta n^{2}_{00}+2\sqrt{4-\delta n^{2}_{00}}\right)\left(2+|\delta n_{00}| \right)}{4 \sqrt{-(-4+\delta n_{00}^2)(4+\delta n_{00}^2+4|\delta n_{00}|)}}\text{.}
\end{align}
The CI coefficients of the non-interacting electrons are shown in the leftmost column of Fig.~\ref{fig:exact-map-alpha-dn}, where $\varphi=0$. Approaching the strictly-localized electron limit, 
i.e. from left to right in Fig.~\ref{fig:exact-map-alpha-dn}, the gradient of the CI coefficients sharpens. 
This sharpening corresponds to the \intra of the $\delta n_{00}$-to-$\delta v$ map introduced in section \ref{sec:features} and is inherited by 
the CI coefficients. Furthermore, the inter-system derivative discontinuity shows up in the CI coefficients for $\varphi=\frac{\pi}{2}$ and 
the CI functionals become distributional points, 
\[
   \alpha_1^{\varphi=\frac{\pi}{2}}[\delta n_{00}] = 
\begin{cases}
\frac{1}{2}  , & \text{for } \delta n_{00}=-2\\
\frac{1}{\sqrt{2}}   , & \text{for } \delta n_{00}= 0\\
\frac{1}{2} , & \text{for } \delta n_{00}=+2 \text{,}
\end{cases}
\]

\[
   \alpha_2^{\varphi=\frac{\pi}{2}}[\delta n_{00}] = 
\begin{cases}
\frac{1}{\sqrt{2}}  , & \text{for } \delta n_{00}=-2\\
0   , & \text{for } \delta n_{00}= 0\\
-\frac{1}{\sqrt{2}} , & \text{for } \delta n_{00}=+2 \text{,}
\end{cases}
\]

\[
   \alpha_3^{\varphi=\frac{\pi}{2}}[\delta n_{00}] = 
\begin{cases}
-\frac{1}{2}  , & \text{for } \delta n_{00}=-2\\
\frac{1}{\sqrt{2}}   , & \text{for } \delta n_{00}= 0\\
-\frac{1}{2} , & \text{for } \delta n_{00}=+2 \text{,}
\end{cases}
\]
which are connected via straight lines due to the degeneracy of the ground-state.
\subsection{Exact Ground-State and Excited-State Energy Functionals}
\label{subsec:hk_func}
\begin{figure*}[ht]
\includegraphics[width=\textwidth,natwidth=6.69in,natheight=5.13in]{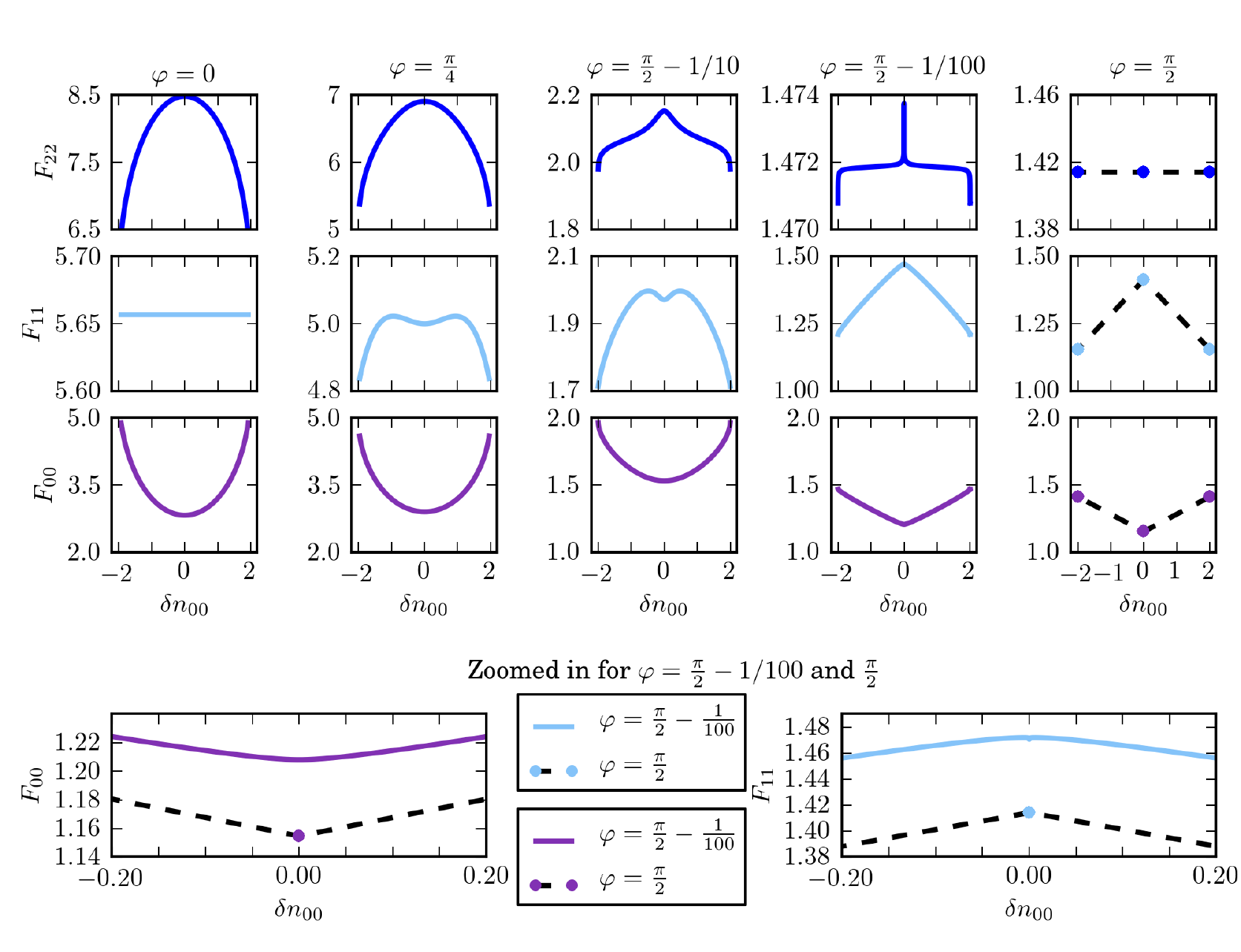}
\caption{Exact energy functionals $F_{jj}=\bra{\Psi_j}\lambda_t(\varphi)\hat{T}+\lambda_w(\varphi)\hat{W}\ket{\Psi_j}$ of the ground-, the first- and second-excited state for different strengths of the electron localization $\varphi$. 
First row: second excited-state energy $F_{22}$ as functional of the ground-state density. 
 Second row: first excited-state energy $F_{11}$ as functional of 
the ground-state density. Third row: ground-state energy $F_{00}$ as functional of the ground-state density, i.e. the Hohenberg-Kohn functional. From the non-interacting limit (left) to the strictly-localized limit (right), the gradient of all energy functionals steepens. 
In the highly localized limit, where $\varphi=\frac{\pi}{2}-\frac{1}{100}$, all energy functionals are continuous. In particular, the ground-state energy functional shows a convex behavior as can be seen in the detailed view of the \intra of highly-localized electrons and the inter-system derivative discontinuity of strictly-localized electrons at the bottom of the figure. Note, that here the $x$-axis has been scaled by one order of magnitude. In the strictly-localized limit, for all energy functionals only the three distributional points $\delta n_{00}= \pm 2$ and $\delta n_{00}=0$ exist. Due to the degeneracy of the eigenstates in the strictly localized limit which is shown in the lower panel of Fig.~\ref{fig:exact-map-dn-dv}, these three distributional points connect via straight lines indicated by a black-dashed line.}
\label{fig:exact_map_e_dn}
\end{figure*}
Since the CI coefficients $\alpha^{\varphi}_q$ of the wavefunction inherit 
the \intra and the inter-system derivative discontinuity, arbitrary ground-state expectation values,
defined in Eq. \ref{eq:observables_ci}, also inherit the 
\intra and the inter-system derivative discontinuity. Note, the excited-state CI coefficients also show the same exact features, which are then inherited by excited-state functionals in the respective limit.
As particular examples for this inheritance, we illustrate in Fig.~\ref{fig:exact_map_e_dn} the \intra and the inter-system derivative discontinuity for the exact Hohenberg-Kohn 
functional ($j=0$) and the excited-state energy functionals  ($j=1,2$) 
\begin{align}
F_{jj}^\varphi[\delta n_{00}] =\bra{\Psi_{\text{2s},j}^{\varphi}}\lambda_{t}(\varphi)\hat{T}+\lambda_{w}(\varphi)\hat{W}\ket{\Psi_{\text{2s},j}^{\varphi}}\text{,}
\label{eq:hk_all_ci}
\end{align}
for the two-particle singlet states $\ket{\Psi_{\text{2s},j}^{\varphi}} = \ket{\Psi_j^\varphi[\delta n_{00}, N=2, S^2=0, S_z=0]}$.
The third row of Fig.~\ref{fig:exact_map_e_dn} shows the exact Hohenberg-Kohn functional ($j=0$) discussed previously in literature \cite{Capelle2003, Fuks2013, Carrascal2012, Carrascal2015}, the first and second row show the first and second excited-state energy functional ($j=1,2$), respectively. The gradient of all three functionals $F_{jj}[\delta n_{00}]$ steepens approaching the limit of strictly localized electrons, just as previously observed for the density-to-potential map in Sec. \ref{sec:features} and the density-to-wavefunction map in Sec. \ref{subsec:ci_coefficients}. However, if $\varphi$ differs infinitesimally from the strictly localized limit, all energy functionals are continuous. In particular, the ground-state energy functional $F_{00}$ is convex. The difference between the highly localized and the strictly localized limit, is displayed in an inset at the bottom in Fig.~\ref{fig:exact_map_e_dn}, which contains a zoom of the critical region of the ground- and first excited-state state functional. Again, in the limit of strictly localized electrons, 
the \intra transitions into the inter-system derivative discontinuity. As already discussed for the density-to-wavefunction map, the distributional points can be connected via straight lines due to the degeneracy of the eigenstates in the strictly localized limit.
\subsection{Exact Excited- and Transition Density Functionals}
\begin{figure*}[ht]
\includegraphics[width=\textwidth,natwidth=6.69in,natheight=4.13in]{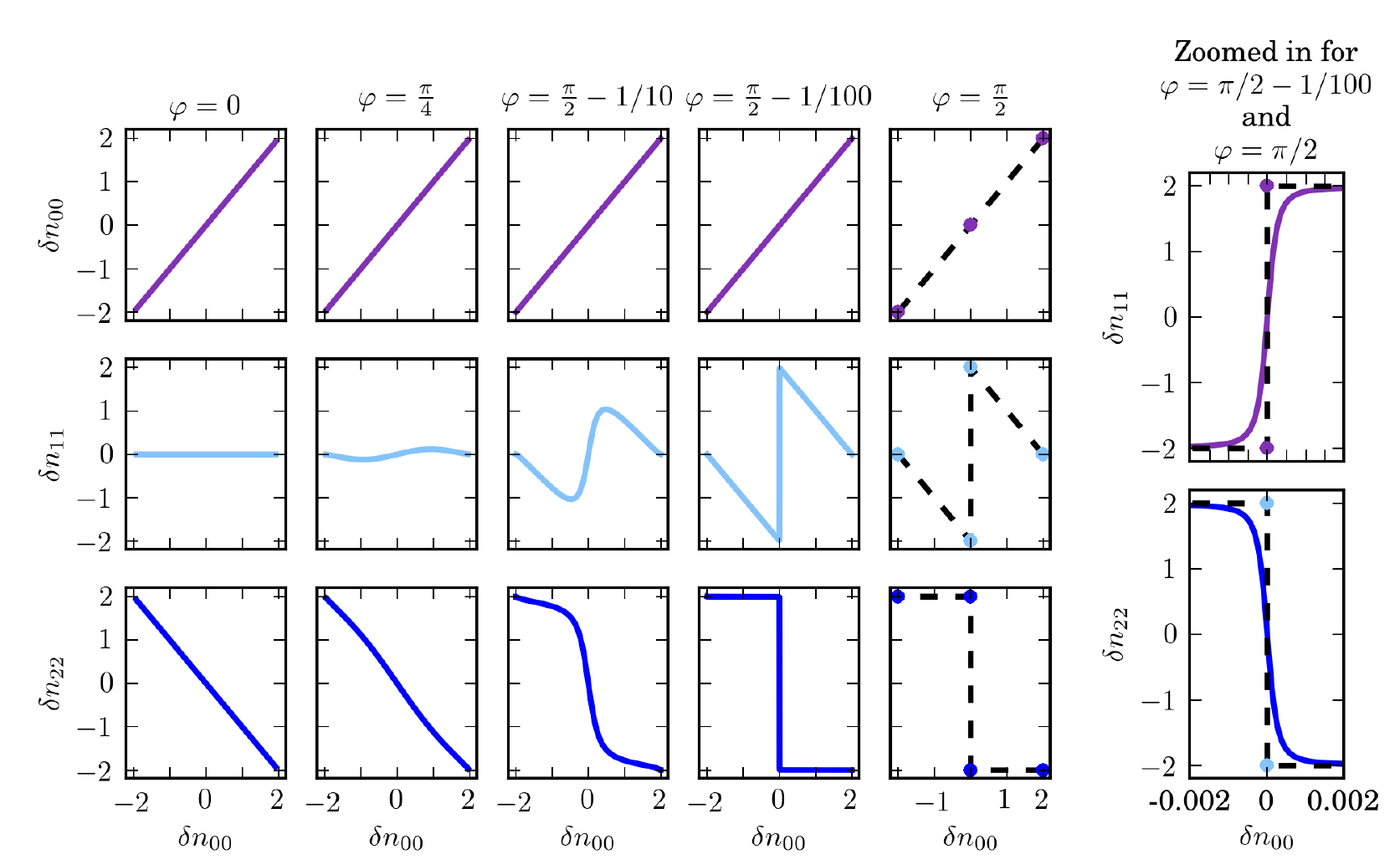}
\caption{Density functionals for ground- and excited-singlet states. First panel: ground-state density as functional of the ground-state density. Second panel: first excited-state density as functional of the ground-state density. Third panel: second excited-state density as functional of the ground-state density. From the non-interacting limit (left) to the strictly-localized limit (right), the gradient of all excited-state density functionals steepens. A detailed view of the \intra for highly-localized electrons ($\varphi=\frac{\pi}{2}-\frac{1}{100}$) and the inter-system derivative discontinuity is given on the right. }
\label{fig:exact-map-dn-dn}
\end{figure*}
\begin{figure*}[ht]
\includegraphics[width=\textwidth,natwidth=6.69in,natheight=3.72in]{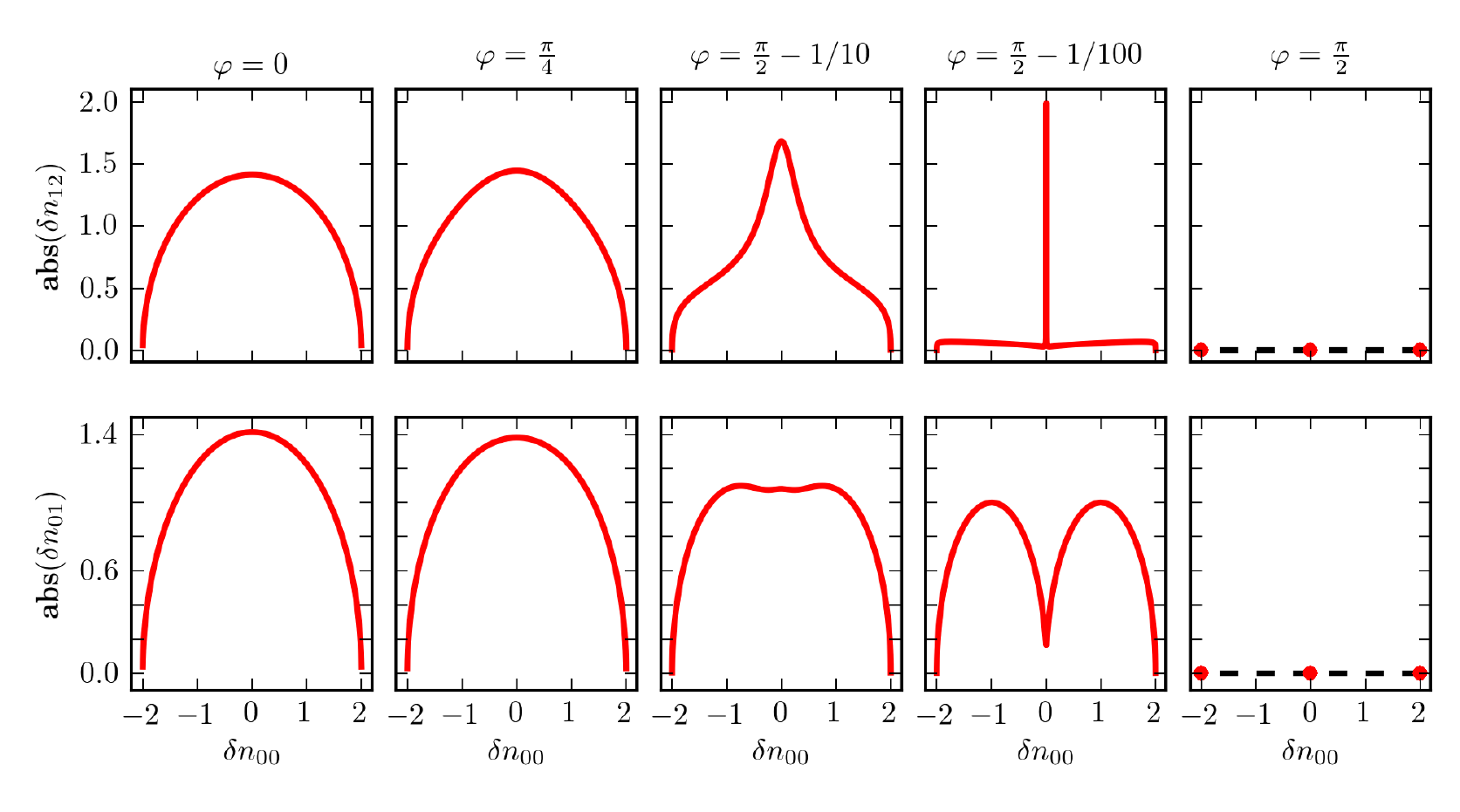}
\caption{Transition matrix elements of the density operator between different excited many-body states as functional of the ground-state density $\delta n_{jk}=\bra{\Psi_j}\delta \hat{n}\ket{\Psi_k}$. First row: Absolute value of the exact transition density from the first and second excited-state as functional of the ground-state density $\delta n_{12}(\delta n_{00})$. Second row: Absolute value of the exact transition density from the ground-and the first excited-state as functional of the ground-state density $\delta n_{01}(\delta n_{00})$. From the non-interacting (left) to the strictly localized limit (right), approaching the strictly localized limit the gradient of both transition density functionals steepens. In the strictly localized limit, the sites are disconnected. Therefore, there are no transitions between the three two-particle singlet states, and the transition densities are zero.}
\label{fig:exact-map-dntrans-dn}
\end{figure*}
To illustrate the fact that all observables inherit the \intra and the inter-system derivative discontinuity,
 we also show the excited- ($k=j=1,2$) and transition-state densities($k\neq j=0,1,2$) 
\begin{equation}
\delta n_{kj} [\delta n_{00}] =\bra{\Psi_k[\delta n_{00}, N]}\hat{O}\ket{\Psi_j[\delta n_{00}, N]}
\end{equation}
as functionals of the ground-state density $\delta n_{00}$.
The excited-state density functionals are shown in the second and third row of Fig.~\ref{fig:exact-map-dn-dn} respectively. For completeness, also the trivial linear behavior of the ground-state density as functional of the ground-state density is shown in the first row of the figure.
From the non-interacting (left) to the strictly-localized limit (right), the gradient of the excited-state density functionals steepens up to the strictly-localized limit where the excited-state density functionals obey the straight-line condition due to the degeneracy of the ground-state. To highlight the difference of the \intra and inter-system derivative discontinuity of the excited-state density functionals a detailed view of the critical region can be found on the right-hand side of Fig.~\ref{fig:exact-map-dn-dn}.\\
Transition densities are an important ingredient for linear response calculations in time-dependent DFT (TDDFT). In TDDFT, the transition densities are often approximated by the ones computed from Kohn-Sham determinants. For our model system, we show the exact transition densities as functionals of the ground-state density.
In contrast to the excited-state density functionals, the transition density functionals are phase-dependent. Fig.~\ref{fig:exact-map-dntrans-dn} shows the absolute value of the transition density as functional of the ground-state density. The first and second row of Fig.~\ref{fig:exact-map-dntrans-dn} show the absolute value of the transition density from the first to the second and from the ground- to the second excited state, respectively. Approaching the strictly localized limit, both transition-state densities show clearly the \intraspace. In the strictly localized limit, there is no transition between the eigenstates of the system and the transition-state densities are zero, see $\varphi=\frac{\pi}{2}$ in panel one and two. 
\subsection{Exact Correlation Entropy Functional}
\begin{figure*}[ht]
\includegraphics[width=\textwidth,natwidth=6.69in,natheight=1.86in]{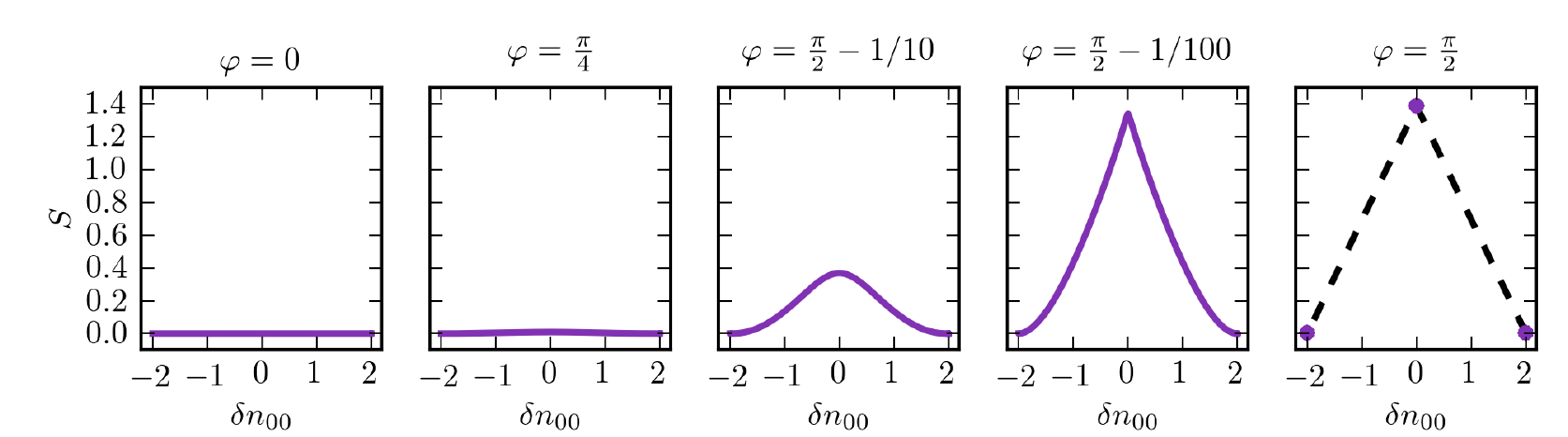}
\caption{Correlation entropy as functional of the ground-state density indicating the correlation within the system. For non-interacting electrons the correlation entropy is zero. From left to right, approaching the strictly localized limit, the correlation and the mixing of the eigenstates and hence the correlation entropy increases. Furthermore, the gradient of the functional obeys the \intra and the inter-system derivative discontinuity for $\varphi=\frac{\pi}{2}$.}
\label{fig:vN}
\end{figure*}
As final example we illustrate the functional behavior of the correlation entropy. The correlation entropy , discussed in detail in Ref.~\cite{Ziesche1997} measures the correlation and entanglement present in a many-body system.
It can be understood as well as a measure of the Slater rank \cite{Ziesche1997, Cirac2001} as can be seen if we compare the correlation entropy plotted in Fig.~\ref{fig:vN} with the mixing of the eigenstates in lower panel and inset of Fig.~\ref{fig:exact-map-dn-dv} for the different values of the parameter $\varphi$.
In the two-site model, where we have access to all eigenvectors and eigenvalues, we can compute the correlation entropy of the system,
\begin{align}
S=\sum_{j=1}^{\infty}n_j \text{ln} n_j \text{,}
\end{align}
where $n_j$ are the eigenvalues of the reduced one-body density matrix
\begin{align}
\rho_{00}(j\sigma,j'\sigma')=\bra{\Psi_0}\hat{c}^{\dagger}_{j\sigma}\hat{c}_{j'\sigma'}\ket{\Psi_0}\text{.}
\end{align}
The correlation entropy is zero for pure states, and has its maximum for maximally mixed states \cite{Ziesche1997, Cirac2001, Franco2013}. In Fig.~\ref{fig:vN} we see that the correlation entropy increases with increasing correlation while the gradient of the correlation entropy functional obeys the \intra and transitions into the inter-system derivative discontinuity for $\varphi=\frac{\pi}{2}$. In the limit of non-interacting electrons, where there is no correlation, the correlation entropy vanishes. The maximum value of the correlation entropy is reached in the strictly localized limit for $\delta n_{00} =0$ where all three eigenenergies are degenerate.
\section{Summary}
\label{sec:summary}
In the present work we have illustrated how the \intraspace, an exact feature of the ground-state density-to-potential map, develops gradually with increasing decoupling between fragments of a system and transforms into the well-known inter-system derivative discontinuity for fully decoupled systems. As a consequence of the Hohenberg-Kohn theorem, the wavefunction-to-density map inherits the exact features of the density-to-potential map. Furthermore, the exact features of the density-to-potential map transmit to ground- and excited-state observables and transition-matrix elements. We illustrated the inheritance of these features by showing the ground- and excited-state energy, the excited- and transition-state densities and the correlation entropy ground-state density functionals.\\ \\
Although both exact features are linked to the localization of the electrons, we carved out that the \intra and the inter-system derivative discontinuity are conceptually different features within density functional theory. The inter-system derivative discontinuity corresponds to the electron localization in {\it fully} decoupled systems with fixed particle number. In the decoupled limit, the Hohenberg-Kohn theorem is not applicable by construction, and the one-to-one density-to-potential map breaks down. The intersystem derivative discontinuity coincides with a real crossing of the eigenenergies of the system as function of the external potential. Ground-state density functionals in the decoupled limit are straight lines between different values for the particle number $N$ due to mixture of states in degenerate subspaces, $F=(1-\omega)F_{N}+\omega F_{N+1}$ with the mixing parameter $0\leq\omega\leq1$. The \intra instead corresponds to the electron localization in coupled fragments of a system, where one fragment can be seen as the particle reservoir (bath) of the other, but the particle number of the total system is fixed. The \intra coincides with an avoided crossing of the eigenenergies as function of the external potential and sharpens when approaching the real crossing. Ground-state density functionals result directly from the one-to-one correspondence of the Hohenberg-Kohn theorem, such as the convex ground-state energy as function of the density difference between the fragments of the system.\\ \\
The inter-system derivative discontinuity plays a crucial role whenever the particle number of the total system changes which is the case for observables such as the electron affinity $A=E[N]-E[N+1]$, the ionization energy $I=E[N-1]-E[N]$, the fundamental gap which is the difference of ionization energy and affinity $E_{gap}=I-A$, and the chemical hardness $\eta=\left( \frac{\partial^2 E}{\partial N^2}\right)_{v}$ of a system. The \intra is linked to processes where particles are transferred from one fragment to another within a system of fixed particle number such as stretched molecules, charge-transfer processes and any problem involving highly-localized electrons. Approximate functionals fail to describe such problems not due to the lack of the inter-system derivative discontinuity but due to the lack of the \intraspace. Given the relevance of the above mentioned problems it is crucial to develop improved density functionals that capture this exact condition of the exact density-to-potential and density-to-wavefunction maps. In the highly localized electron limit the exact xc-functional does not present a straight line behavior as in $E(N)$ but rather a sharp but differentiable one as in E$(\delta n)$, where $\delta n$ represents the density difference between the fragments.\\ \\
Our work illustrates those fundamental concepts of density functional theory. To improve the accuracy of DFT observables, approximate functionals should capture both, the inter-system derivative discontinuity and the \intra respectively. Work about how to generalize the present results from lattice Hamiltonians to real continuous systems is currently in progress.

Our results also allow to get insight about spin DFT functionals as the magnetization of the N electron system can be written in terms of the ground-state density (as all other observables we discussed in this paper). This is a way to solve the known problems of spin DFT \cite{Eschrig2001, Capelle2001} (however it would require going beyond present adiabatic functionals, work along those lines is in progress).
\begin{acknowledgments}
The authors thank Professor Matthias Scheffler for his support, Johannes Flick, Jessica Walkenhorst and Viktor Atalla for very useful discussions and comments, and Nicola Kleppmann, Teresa Reinhard and Anne Hodgson for comments on the manuscript. We acknowledge financial support from the European Research Council Advanced Grant DYNamo (ERC-2010- AdG-267374), Spanish Grant (FIS2013-46159-C3-1-P), Grupos Consolidados UPV/EHU del Gobierno Vasco (IT578-13), COST Actions CM1204 (XLIC), and MP1306 (EUSpec).
\end{acknowledgments}

%\bibliographystyle{aipnum4-1}
%\bibliography{aipsamp}
\bibliography{references}

\end{document}